\documentclass[journal,twocolumn,twoside]{IEEEtran}

\usepackage[utf8x]{inputenc}

\usepackage{amsfonts}
\usepackage{amsmath}
\usepackage{amssymb}

\usepackage{graphicx}

\usepackage{algorithmic}
\renewcommand{\algorithmiccomment}[1]{//#1}

\usepackage{framed}

\usepackage{fixltx2e}
\MakeRobust{\overrightarrow}
\newcommand{\grad}{\overrightarrow\bigtriangledown}

\newcommand{\dx}{\partial_{x}}
\newcommand{\dy}{\partial_{y}}
\newcommand{\dz}{\partial_{z}}
\newcommand{\dc}{\partial_{c}}

\title{Jacobians and Hessians of Mean Value Coordinates for Closed Triangular
Meshes}
\author{Jean-Marc Thiery %
\hspace{0.5cm} Julien Tierny%
\hspace{0.5cm} Tamy Boubekeur\\ T\'el\'ecom ParisTech -- CNRS -- LTCI}

\begin{document}

\maketitle

\begin{abstract}
In this technical note, we present the formulae of the derivatives of the Mean 
Value Coordinates~\cite{ju2005mean} based transformations
using an enclosing triangle mesh, acting as a cage for the deformation of an
interior object.
\end{abstract}

\section{Background}

\subsection{Mean Value Coordinates for Closed Triangular Meshes}

Mean Value Coordinates for closed triangular meshes were introduced 
in~\cite{ju2005mean}. In this section, we briefly review this work and describe
the notations we will use in the rest of this note.

As written in \cite{ju2005mean}, a 3D point $\eta$ can be expressed as a linear sum of the 3D positions $p_i$ of the vertices of a triangular mesh $M$ by:
$\eta = \frac{\sum_i{w_i \cdot p_i}}{\sum_i{w_i}}  = \sum_i{\lambda_i \cdot p_i}$.

For a point $x$ onto the surface (a two-dimensional parameter), we note as usual $\phi_i[x]$ the linear function on $M$ that takes value 1 on vertex $i$ and 0 on other vertices,
and $p[x]$ its 3D position.
The definition of the weights $\lambda_i$ should guarantee \textit{linear 
precision} (i.e. $\eta = \sum_i{\lambda_i(\eta) p_i}$).

Since $\int_{B_\eta(M)}{\frac{p[x] - \eta}{|p[x] - \eta|}dS_\eta(x)} = 0$ (the integral of the unit outward normal onto the unit sphere is $0$),
we have 
\begin{equation}
\label{eq:linear_precision}
 \eta = \frac{ \int_{B_\eta(M)}{\frac{p[x]}{|p[x] - \eta|}dS_\eta(x)} }{ \int_{B_\eta(M)}{\frac{1}{|p[x] - \eta|}dS_\eta(x)} }
\end{equation}
$B_\eta(M)$ being the projection of the manifold $M$ onto the unit sphere centered in $\eta$.

Writing that $\forall x~p[x] = \sum_i{\phi_i[x] p_i}$, with $\sum_i{
\phi_i[x] } = 1$, we obtain

\begin{equation}
 \eta = \frac{ \sum_i{ \int_{B_\eta(M)}{\frac{\phi_i[x]}{|p[x] - \eta|}dS_\eta(x)} p_i} }{ \int_{B_\eta(M)}{\frac{1}{|p[x] - \eta|}dS_\eta(x)} }
\end{equation}

The weights $\lambda_i$ are given by
\begin{equation}
\label{eq:lambda_i_on_M}
 \lambda_i = \frac{ \int_{B_\eta(M)}{\frac{\phi_i[x]}{|p[x]-\eta|} dS_\eta(x)} }{ \int_{B_\eta(M)}{\frac{1}{|p[x] - \eta|}dS_\eta(x)} }
\end{equation}

And the weights $w_i$ such that $\lambda_i = \frac{w_i}{\sum_j{w_j}}$ are given by
\begin{equation}
\label{eq:w_i_on_M}
 w_i = \int_{B_\eta(M)}{\frac{\phi_i[x]}{|p[x]-\eta|} dS_\eta(x)}
\end{equation}

This definition guarantees linear precision; it gives a linear interpolation of 
the function onto the triangles of the cage; and it
extends it in a regular way to the entire 3D space.

\paragraph*{Computing the weights $w_i$}
The support of the function $\phi_i[x]$ is only composed of 
the adjacent triangles to the vertex $i$. Then,  we can rewrite 
Eq.~\ref{eq:w_i_on_M}
as $w_i = \sum_{T \in N1(i)}{w_{i}^{T}}$, with

\begin{equation}
\label{eq:w_i_T_on_T}
w_{i}^{T} = \int_{B_\eta(T)}{\frac{\phi_i[x]}{|p[x]-\eta|} \cdot d\overline{T}}
\end{equation}

Given a triangle $T$ with vertices $t_1, t_2, t_3$, we see that
\begin{equation}
\begin{split}
\sum_{j}{w_{t_j}^{T} \cdot (p_{t_j} - \eta)} =& \int_{B_\eta(T)}{\frac{\sum_{j}{\phi_{t_j}[x] \cdot (p_{t_j} - \eta)}}{|p[x]-\eta|} \cdot d\overline{T}} \\
	=& \int_{B_\eta(T)}{\frac{p[x]-\eta}{|p[x]-\eta|} \cdot d\overline{T}} \triangleq m^{T}
\end{split}
\end{equation}

This last integral is simply the integral of the unit outward normal on the spherical triangle $\overline{T}$.

\begin{figure}[b]
\begin{center}
\includegraphics[width=0.7\linewidth]{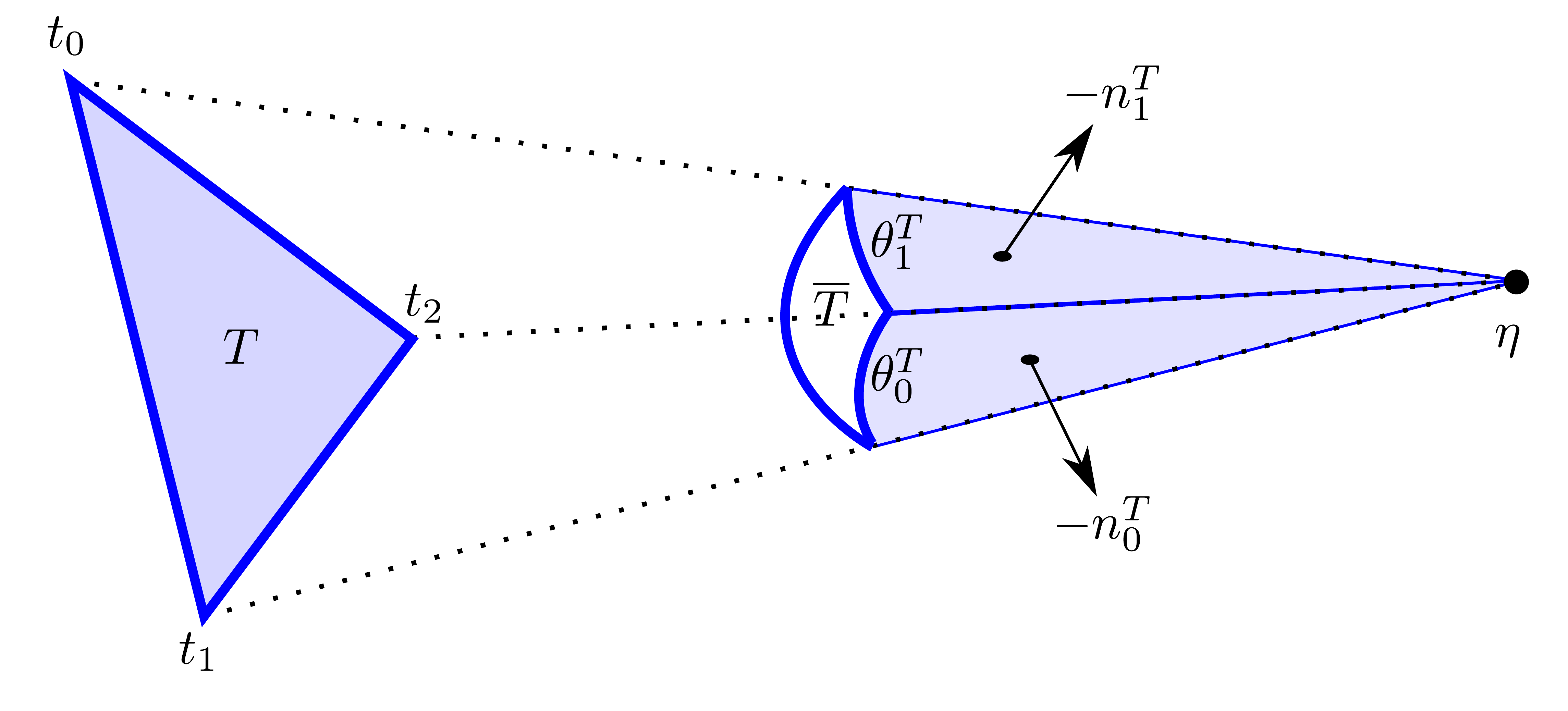}
\end{center}
\caption{Triangle T projected on the spherical triangle \=T.}
\label{fig:spherical_triangle}
\end{figure}

By noting $n_i^T = \frac{N_i^T}{|N_i^T|}$, with $N_i^T \triangleq (p_{t_{i+1}}-\eta) \wedge (p_{t_{i+2}}-\eta)$
(see Fig.\ref{fig:spherical_triangle}), it can be easily expressed as

\begin{equation}
\label{eq:m_T}
 m^{T} = \sum_{i}{ \frac{1}{2} \theta_i^T n_i^T }
\end{equation}

This comes from the fact that the integral of the unit outward normal on a closed surface is always 0.

Finally, we obtain
\begin{equation}
\label{eq:main_equation_of_MVC}
 \sum_{j}{w_{t_j}^{T} \cdot (p_{t_j} - \eta)} = m^{T}
\end{equation}

This point was discussed in \cite{ju2005mean}. As the authors pointed out, 
by noting $A^{T}$ the 3 by 3 matrix $\{p_{t_1}-\eta,p_{t_2}-\eta,p_{t_3}-\eta\}$,
 we can derive the weights $w_{t_j}^{T}$ by

\begin{equation}
\label{eq:weights_by_A_inversion}
 \{w_{t_1}^{T},w_{t_2}^{T},w_{t_3}^{T}\}^t = {A^T}^{-1} \cdot m^{T}
\end{equation}

Since ${N_i^T}^t \cdot (p_{t_j} - \eta) = 0 ~~ \forall i \neq j$, we have from Eq.~\ref{eq:main_equation_of_MVC} that
\begin{equation}
w_{t_i}^{T} = \frac{ {N_i^T}^t \cdot m^T }{ {N_i^T}^t \cdot (p_{t_i} - \eta) }
= \frac{ {N_i^T}^t \cdot m^T }{\det(A^T)}~~\forall \eta \notin Support(T)
\end{equation}

\section{MVC Derivatives}

We now present the derivatives of the Mean Value Coordinates.
Deforming the cage mesh with $f(p_i) = \overline{p_i}$ induces a deformation of 
the 3D space by
$f = \sum_i{\lambda_i \cdot \overline{p_i}}$.
In the rest of the document, for any function $h: E \rightarrow F$, we note $\dx h, \dy h, \dz h$
its derivative by $x, y,$ and $z$, $\grad{h}$ its gradient, $Jh$ its jacobian, and $Hh$ its hessian.
\newline

The deformation function $f$ as defined acts now on $\mathbb{R}^3$ entirely.
The derivatives of $f$
can be expressed as a linear sum of positions $\overline{p_i} = \{\overline{x_i},\overline{y_i},\overline{z_i}\}$:

\begin{equation}
\left\{
    \begin{array}{ll}
Jf &= \sum_i{\overline{p_i} \cdot \grad{\lambda_i}^{t}} \\
H(f_x) &= \sum_i{\overline{x_i} \cdot H\lambda_i} \\
H(f_y) &= \sum_i{\overline{y_i} \cdot H\lambda_i} \\
H(f_z) &= \sum_i{\overline{z_i} \cdot H\lambda_i} \\
    \end{array}
\right.
\end{equation}

Consequently, it allows to specify implicit equations on the cage
in a linear system by giving specified rotations and scales
on 3D locations, or to minimize the norm of the hessian to force rigidity,
as done in the case of Green Coordinates in~\cite{ben2009variational}.

Since $\lambda_i = \frac{ w_i }{ \sum_j{ w_j } }$,
\begin{equation}
 \label{eq:grad_lamda_i}
 \grad{\lambda_i} = \frac{\grad{w_i}}{\sum_j{w_j}} - \frac{ w_i \cdot \sum_j{ \grad{w_j} } }{ (\sum_j{w_j})^2 }
\end{equation}

We also have
$\forall c = x,~y,~z$
\begin{equation}
\begin{split}
 \dc(\grad{\lambda_i}) = &\frac{\dc(\grad{w_i})}{\sum_j{w_j}} - \frac{ \grad{w_i} \cdot \sum_j{\dc(w_j)} }{ (\sum_j{w_j})^2 } \\
			    &- \frac{ \dc(w_i) \cdot \sum_j{\grad{w_j}} + w_i \cdot \sum_j{\dc(\grad{w_j})} }{ (\sum_j{w_j})^2 } \\
			    &+ \frac{ 2 w_i \cdot (\sum_j{\grad{w_j}}) \cdot (\sum_k{\dc(w_k)}) }{ (\sum_j{w_j})^3 }
\end{split}
\end{equation}

or
\begin{equation}
\label{eq:H_lambda_i}
\begin{split}
 H\lambda_i = &\frac{Hw_i}{\sum_j{w_j}} - \frac{ w_i \sum_j{Hw_j} }{ (\sum_j{w_j})^2 } \\
	      & - \frac{ \grad{w_i} \cdot \sum_j{\grad(w_j)^t}  +  \sum_j{\grad(w_j) \cdot \grad{w_i}^t} }{ (\sum_j{w_j})^2 } \\
	      & + \frac{ 2 w_i  (\sum_j{\grad{w_j}}) \cdot (\sum_j{\grad{w_j}})^t }{ (\sum_j{w_j})^3 }
\end{split}
\end{equation}

From these expressions, we see that, in order to get $\grad{\lambda_i}(\eta)$ and $H\lambda_i(\eta)$, we first need to obtain
$\grad{w_i}(\eta)$ and $Hw_i(\eta)$ for each vertex $i$ of the cage.

\subsection*{Special case: $\eta$ lies on the surface of the cage}

Mean Value Coordinates define an \textit{interpolation} process. The function 
represented onto the vertices of the cage
(in our case, a space transformation) is extended to the interior of the
triangles with linear interpolation on each triangle.
Then it is extended to the space by means of a surfacic integration of the 
function (see Eq.~\ref{eq:linear_precision}).

Since we represent the cage as triangle mesh in the 3D case, \textbf{the deformation function cannot
be anything more than continuous onto the edges of the cage in 3D}.
Therefore Jacobians and Hessians of the deformation cannot be evaluated everywhere on the surface of the cage,
and we do not provide any formula for Jacobians and Hessians of the deformation onto the surface of the cage.

\subsection{Expression of the Jacobians}

In the general case where $\det(A^T) = (p_{e_i}-\eta)^t \cdot N_{i}^T \neq 0$, we have

$\grad{w_{t_i}^T} = \frac{{B^T}^t \cdot N_i^T}{\det(A^T)}$

with
\begin{equation*}
\begin{split}
 B^T =& \sum_j{ \frac{eq_1(\theta_j^T) N_j^T \cdot {N_j^T}^t \cdot JN_j^T}{2 (|p_{t_{j+2}}-\eta| |p_{t_{j+1}}-\eta|)^3} } \\
	&- \sum_j{ \frac{N_j^T \cdot (2 \eta - p_{t_{j+1}} - p_{t_{j+2}})^t}{2 (|p_{t_{j+2}}-\eta| |p_{t_{j+1}}-\eta|)^2} } \\
	&+ \sum_j{ \frac{eq_2(\theta_j^T) JN_j^T}{2 |p_{t_{j+2}}-\eta| |p_{t_{j+1}}-\eta|} }  + \sum_j{ w_{t_j}^T \cdot I_3 }
\end{split}
\end{equation*}

and $eq_1(x) = \frac{\cos(x)\sin(x)-x}{\sin(x)^3}$ and $eq_2(x)=\frac{x}{\sin(x)}$ two well defined functions on $]0,\pi[$
that admit well controlled Taylor expansion around $0$, 
$JN_j^T = (p_{t_{j+2}} - p_{t_{j+1}})_{[\wedge]}$,
$k_{[\wedge]}$ being the skew 3 by 3 matrix (i.e. ${k_{[\wedge]}}^t = -k_{[\wedge]}$)
such that $k_{[\wedge]} \cdot u = k \wedge u ~~ \forall k,u \in \mathbb{R}^3$.

\subsubsection*{Special case: $\eta \in Support(T), \notin T$ }
\label{sec:grad_special_case}

\begin{equation}
\begin{split}
-2 |T| \grad{w_i^T} =& \sum_j{\frac{eq_2(\theta_j^T) (p_{t_{i+2}} - p_{t_{i+1}})^t \cdot (p_{t_{j+2}} - p_{t_{j+1}})}{2 |p_{t_{j+2}}-\eta||p_{t_{j+1}}-\eta|}} n_T \\
   &+ \sum_j{\frac{eq_1(\theta_j^T)|p_{t_{j+2}}-p_{t_{j+1}}|^2 {N_i^T}^t \cdot N_j^T}{4 (|p_{t_{j+2}}-\eta||p_{t_{j+1}}-\eta|)^3}} n_T \\
   &+ \sum_j{\frac{\cos(\theta_j^T) eq_3(\theta_j^T) {N_i^T}^t \cdot N_j^T}{2 (|p_{t_{j+2}}-\eta||p_{t_{j+1}}-\eta|)^2}} n_T
\end{split}
\end{equation}

with $eq_2(x) = \frac{x}{\sin(x)},
eq_1(x) = \frac{\cos(x)\sin(x) - x}{\sin(x)^3},$
and $eq_3(x) = \frac{\cos(x) - 1}{\sin(x)^2}$
being functions well defined on $]0 , \pi[$ and that admit controlable Taylor expansion around $0$.

\subsection{Expression of the Hessians}

We note $\delta^x = \left( {\begin{array}{c} 1 \\ 0 \\ 0 \end{array} } \right),\delta^y = \left( {\begin{array}{c} 0 \\ 1 \\ 0 \end{array} } \right),
\delta^z = \left( {\begin{array}{c} 0 \\ 0 \\ 1 \end{array} } \right)$.

\begin{equation}
\begin{split}
 Hw_i^T &= \frac{ 1 }{ \det(A^T) } \left( \begin{array}{ccc} {N_i^T}^t \cdot \dx(Jm^T) \\ {N_i^T}^t \cdot \dy(Jm^T) \\ {N_i^T}^t \cdot \dz(Jm^T) \end{array} \right) \\
	  +& \frac{ 1 }{ \det(A^T) } ( N_i^T \cdot (\sum_j{\grad{w_j^T}})^t  +  \sum_j{\grad{w_j^T}} \cdot {N_i^T}^t )
\end{split}
\end{equation}

with
\begin{equation}
\begin{split}
 & \dc(Jm^T) = \sum_j{ \frac{eq_6(\theta_j^T) ({JN_j^T}^t\cdot N_j^T)_{(c)} N_j^T \cdot {N_j^T}^t \cdot JN_j^T}{2 (|p_{t_{j+2}}-\eta| |p_{t_{j+1}}-\eta|)^5} } \\
	&- \sum_j{ \frac{eq_7(\theta_j^T) (2 \eta - p_{t_{j+1}} - p_{t_{j+2}})_{(c)} N_j^T \cdot {N_j^T}^t \cdot JN_j^T}{2 (|p_{t_{j+2}}-\eta| |p_{t_{j+1}}-\eta|)^4} } \\
	&+ \sum_j{ \frac{eq_1(\theta_j^T) \dc(N_j^T) \cdot {N_j^T}^t \cdot JN_j^T}{2 (|p_{t_{j+2}}-\eta| |p_{t_{j+1}}-\eta|)^3} } \\
	&+ \sum_j{ \frac{eq_1(\theta_j^T) N_j^T \cdot {\dc(N_j^T)}^t \cdot JN_j^T}{2 (|p_{t_{j+2}}-\eta| |p_{t_{j+1}}-\eta|)^3} } \\
	&- \sum_j{ \frac{3 eq_1(\theta_j^T) (\eta - p_{t_{j+1}})_{(c)} N_j^T \cdot {N_j^T}^t \cdot JN_j^T}{2 |p_{t_{j+2}}-\eta|^3 |p_{t_{j+1}}-\eta|^5} } \\
	&- \sum_j{ \frac{3 eq_1(\theta_j^T) (\eta - p_{t_{j+2}})_{(c)} N_j^T \cdot {N_j^T}^t \cdot JN_j^T}{2 |p_{t_{j+2}}-\eta|^5 |p_{t_{j+1}}-\eta|^3} } \\
	&- \sum_j{ \frac{\dc(N_j^T) \cdot (2 \eta - p_{t_{j+1}} - p_{t_{j+2}})^t}{2 (|p_{t_{j+2}}-\eta| |p_{t_{j+1}}-\eta|)^2} } \\
	&+ \sum_j{ \frac{(\eta - p_{t_{j+1}})_{(c)} N_j^T \cdot (2 \eta - p_{t_{j+1}} - p_{t_{j+2}})^t}{ |p_{t_{j+2}}-\eta|^2 |p_{t_{j+1}}-\eta|^4} } \\
	&+ \sum_j{ \frac{(\eta - p_{t_{j+2}})_{(c)} N_j^T \cdot (2 \eta - p_{t_{j+1}} - p_{t_{j+2}})^t}{ |p_{t_{j+2}}-\eta|^4 |p_{t_{j+1}}-\eta|^2} } \\
	&+ \sum_j{ \frac{eq_8(\theta_j^T) ({JN_j^T}^t\cdot N_j^T)_{(c)} JN_j^T}{2 (|p_{t_{j+2}}-\eta| |p_{t_{j+1}}-\eta|)^3} } \\
	&- \sum_j{ \frac{eq_9(\theta_j^T) (2 \eta - p_{t_{j+1}} - p_{t_{j+2}})_{(c)} JN_j^T}{2 (|p_{t_{j+2}}-\eta| |p_{t_{j+1}}-\eta|)^2} } \\
	&- \sum_j{ \frac{(\eta - p_{t_{j+1}})_{(c)} eq_2(\theta_j^T) JN_j^T}{2 |p_{t_{j+2}}-\eta| |p_{t_{j+1}}-\eta|^3} } \\
	&- \sum_j{ \frac{(\eta - p_{t_{j+2}})_{(c)} eq_2(\theta_j^T) JN_j^T}{2 |p_{t_{j+2}}-\eta|^3 |p_{t_{j+1}}-\eta|} } \\
	&- \sum_j{ \frac{N_j^T \cdot {\delta^c}^t}{|p_{t_{j+2}}-\eta|^2 |p_{t_{j+1}}-\eta|^2} }
\end{split}
\end{equation}

and $eq_6(x) = \frac{d(eq_1)}{dx}(x)cos(x)/sin(x), eq_7(x)=\frac{d(eq_1)}{dx}(x)sin(x), eq_8(x)=\frac{d(eq_2)}{dx}(x)cos(x)/sin(x)$, and $eq_9(x)=\frac{d(eq_2)}{dx}(x)sin(x)$ 
being functions well defined on $]0 , \pi[$ and that admit controllable Taylor expansion around $0$.

\subsubsection*{Special case: $\eta \in Support(T), \notin T$ }

\begin{equation}
H(w_i^T)(\eta) = \grad{dw_i^T}(\eta) \cdot n_T^t
\end{equation}

with

\begin{equation}
\begin{split}
& -2 |T| \grad{dw_i^T} = \\
& - \sum_j{ \frac{((p_{t_{i+2}} - p_{t_{i+1}})^t \cdot (p_{t_{j+2}} - p_{t_{j+1}})) (2 \eta - p_{t_{j+2}} - p_{t_{j+1}})  }{(|p_{t_{j+2}}-\eta||p_{t_{j+1}}-\eta|)^2} } \\
		& + \sum_j{ \frac{eq_1(\theta_j^T) ((p_{t_{i+2}} - p_{t_{i+1}})^t \cdot (p_{t_{j+2}} - p_{t_{j+1}})) {JN_j^T}^t \cdot N_j^T}{2 (|p_{t_{j+2}}-\eta||p_{t_{j+1}}-\eta|)^3} } \\
		& + \sum_j{ \frac{|p_{t_{j+2}}-p_{t_{j+1}}|^2 ({N_i^T}^t \cdot N_j^T) (2 \eta - p_{t_{j+2}} - p_{t_{j+1}})}{2 (|p_{t_{j+2}}-\eta||p_{t_{j+1}}-\eta|)^4} } \\
		& + \sum_j{ \frac{eq_1(\theta_j^T) |p_{t_{j+2}}-p_{t_{j+1}}|^2 ({JN_j^T}^t \cdot N_i^T + {JN_i^T}^t \cdot N_j^T)}{4 (|p_{t_{j+2}}-\eta||p_{t_{j+1}}-\eta|)^3} } \\
		& - \sum_j{ \frac{eq_4(\theta_j^T) |p_{t_{j+2}}-p_{t_{j+1}}|^2 ({N_i^T}^t \cdot N_j^T) {JN_j^T}^t \cdot N_j^T }{2 (|p_{t_{j+2}}-\eta||p_{t_{j+1}}-\eta|)^5} } \\
		& + \sum_j{ \frac{\cos(\theta_j^T) eq_3(\theta_j^T) ({JN_j^T}^t \cdot N_i^T + {JN_i^T}^t \cdot N_j^T) }{2 (|p_{t_{j+2}}-\eta||p_{t_{j+1}}-\eta|)^2} } \\
		& - \sum_j{ \frac{(1 - 2 \cos(\theta_j^T)) ({N_i^T}^t \cdot N_j^T) (2 \eta - p_{t_{j+2}} - p_{t_{j+1}}) }{2 (|p_{t_{j+2}}-\eta||p_{t_{j+1}}-\eta|)^3} } \\
		& + \sum_j{ \frac{eq_5(\theta_j^T) ({N_i^T}^t \cdot N_j^T) {JN_j^T}^t \cdot N_j^T}{2 (|p_{t_{j+2}}-\eta||p_{t_{j+1}}-\eta|)^4} }
\end{split}
\end{equation}

and $eq_1(x)=\frac{\cos(x)\sin(x) - x}{\sin(x)^3},
eq_4(x)=\frac{2 \cos(x)\sin(x)^3 + 3 (\sin(x)\cos(x) - x)}{\sin(x)^5}, 
eq_3(x)=\frac{\cos(x) - 1}{\sin(x)^2},$
and $eq_5(x)=\frac{\cos(x)\sin(x)^2 (1 - 2\cos(x)) - 2 \cos(x)^2 + 2 \cos(x)}{\sin(x)^4}$
being functions well defined on $]0 , \pi[$ and that admit controllable Taylor expansion formula around $0$.

\bibliographystyle{amsplain}
\bibliography{MVCderivatives}

\end{document}